# Increasing Gender Diversity and Inclusion in Scientific Committees and Related Activities at STScI

Gisella De Rosa[1], Cristina Oliveira[1], Camilla Pacifici[1], Alessandra Aloisi[1], Katey Alatalo[1], Trisha Ashley[1], Tracy Beck[1], Martha Boyer[1], Annalisa Calamida[1], Joleen Carlberg[1], Carol Christian[1], Christine Chen[1], Susana Deustua[1], Karoline Gilbert[1], Lea Hagen[1], Alaina Henry[1], Svea Hernandez[1], Bethan James[1], Susan Kassin[1], Stephanie La Massa[1], Jennifer Lotz[1], Margaret Meixner[1], Ivelina Momcheva[1], Amaya Moro-Martin[1], Laura Prichard[1], Swara Ravindranath[1], Julia Roman-Duval[1], Elena Sabbi[1], Elena Sacchi[1], Hannah Wakeford[1], and Tea Temim[1]

[1]Space Telescope Science Institute, 3700 San Martin Drive, Baltimore, MD 21218, USA

**Abstract**

We present a new initiative by the Women in Astronomy Forum at Space Telescope Science Institute (STScI) to increase gender diversity and inclusion in STScI's scientific committees and the activities they generate. This initiative offers new and uniform guidelines on binary gender representation goals for each committee and recommendations on how to achieve them in a homogeneous way, as well as metrics and tools to track progress towards defined goals. While the new guidelines presented in the paper focus on binary gender representation, they can be adapted and implemented to support all minority groups. By creating diverse committees and making them aware of, and trained on implicit bias, we expect to create a diverse outcome in the activities they generate, which, in turn, will advance science further and faster.

**1. Background and Motivation**

Diversity and inclusion are key core values adopted by the Space Telescope Science Institute (STScI), and the institute has taken many steps through the years to improve equality. Past efforts focused on the creation of inclusion, diversity and affinity groups, providing all gender restrooms and lactation rooms, extending medical coverage to pay for transgender transition, same sex/domestic partner medical and dental benefits, offering teleworking options, flexible work schedules, paid parental leave, and mentoring on career advancement. STScI has also instituted a double-anonymous selection process for HST proposals, a first-of-its-kind peer-review process for allocating time on NASA's facilities, that led in 2018 to higher success rate of women-led proposals for the first time in 18 years (Strolger & Natarajan, 2019).

STScI currently has 164 research staff members. A number of research staff-related committees carry out a variety of activities, ranging from recruiting, evaluating, and promoting research staff, to organizing the weekly colloquium and the yearly scientific symposium. Ensuring a balanced gender representation in the STScI's scientific committees has been left to the discretion of the committee's chair, with sometimes less than successful outcomes. To address this issue, the Women in Astronomy Forum (WiAF), one of STScI's affinity groups, created new and uniform guidelines on binary gender representation goals for each committee and recommendations on how to achieve them in a homogeneous way, as well as metrics and tools to track progress towards defined goals. The implementation of these guidelines is still underway at STScI, after getting the full support of STScI's leadership. While it is too early to evaluate the results of this new initiative, we encourage other institutions to adopt similar guidelines, as by creating diverse committees and making them aware of and trained on implicit bias, the related activities of those committees will create a diverse outcome (Casadevall & Handelsman, 2014) that will advance science further and faster.

## 2. Proposed Guidelines

The guidelines proposed by the WiAF cover the full cycle of work of each scientific committee, from its convening to the final deliverables. The guidelines are currently centered on binary gender representation (women/men), with gender being assigned based on name and/or apparent gender expression. While we recognize that gender is non-binary, this information is usually unavailable for privacy reasons. The guidelines are intended to be generic and valid for all scientific committees and can be grouped in three broad categories:

- **Gender Composition:** Every scientific committee needs to ensure appropriate gender representation in its composition and in its deliverables. Care should be taken to properly acknowledge the employee's workload in committees during their performance appraisal. This is particularly important considering that anecdotal evidence suggests that women and minority scientists have typically a higher committee workload.

- **Implicit Bias:** Since awareness is the most effective starting point to reduce implicit bias (Wilson & Brekke 1994; Green et al. 2007), committee members should receive implicit bias training before the start of the committee's activities. Additional training material on implicit bias should also be made available, tailored to the specific activities of the science committees. For further awareness on the topic, it is recommended that all the committee members take the implicit bias tests on Gender-Science, Gender-Career, and Race developed by the [Project Implicit](https://implicit.harvard.edu/implicit/aboutus.html)[1].

---

[1] https://implicit.harvard.edu/implicit/aboutus.html

- **Evaluation and Recognition:** It is important that the chair of each science committee compiles a self-assessment report explaining the results of the committee towards the gender composition guidelines and goals. The outcome of this process should be taken into account during the chair's performance appraisal. In addition, successful committees should be rewarded (e.g. recognized by their peers) in order to establish a virtuous cycle and reinforce the positive results.

It is important to note that this approach does not try to change the outcome of scientific committees activities' by enforcing strict gender quotas, but instead provides methods to achieve improved gender diversity in the committee's deliverables by increasing the diversity in the committee composition, by training committee members on diversity issues, and by rewarding successful committees.

## 3. Definition of Goals and Metrics

The goals and metrics to be applied both to the committees themselves and their deliverables are defined below, following the guidelines outlined above.

### 3.1 Goals and Metrics: Scientific Committees

The goals and metrics concerning the committees per se are given in Table 1. These are generic and apply to all of the scientific committees. Regarding the gender composition, the committees are encouraged to aim at having women representation similar to the composition of the population of graduate students and early career researchers working in astronomy related fields in the US, currently estimated at 40% (*AAS Status, A report on women in Astronomy*, 2016). The women representation in the committee should not be lower than the coeval composition of the science staff, currently equal to 27% at STScI. The metrics for the scientific committees should be evaluated yearly and the results tracked to understand how well the new guidelines are permeating the culture of the science staff. Since the goals only deal with the composition and the organization of the committees themselves, they should be fairly easy to implement, and consequently goals are expected to be reached within 1 year of adoption of the guidelines.

**Table 1:** Goals and metrics for the scientific committees at STScI

| Goals | Metrics |
| --- | --- |
| Gender composition | Women ratio of 40% (27 % floor) |
| Implicit bias training | Performed? [yes/no] |
| Chair self assessment | Provided? [yes/no] |

**3.2 Goals and Metrics: Deliverables of the Scientific Committees**

Similar to other institutions, scientific committees at STScI are responsible for a variety of activities, including hiring of science staff, deliberating on promotions and renewals, evaluating the yearly scientific productivity of the science staff, allocating grants for scientific projects, and organizing scientific meetings at STScI. With such a variety of tasks, the goals and metrics for the committees' deliverables needed to be highly tailored. As an example, Table 2 shows the goals and metrics for the deliverables of the scientific organizing committee (SOC) of scientific meetings (e.g. symposia, workshops) while Table 3 shows the goals and metrics for the deliverables of a committee that evaluates science performance (science evaluation committee; SEC).

The main deliverable of SOCs consists of lists of speakers for invited and contributed talks. As such, the goals consist of having a diverse composition of conference attendees, invited speakers, and accepted contributed talks. For these deliverables the SOC should aim at a women representation corresponding to the composition of the population of graduate students and early career researchers working in astronomy related fields in the US, currently estimated at 40% (*AAS Status, A report on women in Astronomy*, 2016). The women representation should not be lower than the average composition of the astronomy field in the US, currently equal to 30% (including later career stages). It is also recommended that the SOCs build diversity plans during the meeting planning stage that include strategies to meet the gender diversity goal.

**Table 2:** Goals and metrics for the deliverables of the Symposium scientific organizing committee (SOC)

| Goals | Metrics |
|---|---|
| Gender composition of attendees | Women ratio of 40% (30% floor) |
| Gender composition of invited talks | Women ratio of 40% (30% floor) |
| Gender composition of contributed talks | Women ratio of 40% (30% floor) |
| Diversity plan | Developed? [yes/no] |

At STScI, the SEC is in charge of reviewing the science productivity of the science staff on a yearly basis. The evaluation is based on self-assessments compiled by the science staff members and submitted to the Science Mission Office. The SEC members review and evaluate the contributions based on four categories (exceeds expectations, meets expectations, does not meet expectations, did not submit evaluation). For this deliverable it is important to track the gender composition in each of the categories of evaluation, in order to identify any possible bias in the process. This translates into cross-checking that, for each category, the women in that category, normalized to the total number of women in the science staff, is comparable to the men in the same category, normalized to the total number of men in the science staff (Table 3).

The metrics for the deliverables of the scientific committees should be evaluated and tracked yearly. However, it is likely that some of these goals will only be fully reached on 3-5 years timescale, as quotas are not being imposed, but instead a cultural shift within the science personnel needs to happen. Periodic "lessons learned" reviews as well as monitoring of annual results are useful tools to understand trends and perform course corrections, if necessary.

**Table 3:** Goals and metrics for the deliverables of the science evaluation committee (SEC)

| Goals | Metrics |
| --- | --- |
| Balanced gender composition in the categories of evaluation | Women ratio normalized to women population comparable to men ratio normalized to men population |

### 4. Tracking Results and Identifying Success

In order to assess the effectiveness of the new guidelines it is important to track the metrics through time and compare the results with a baseline obtained before the adoption of the gender guidelines. To do this at STScI, we created a database that collects all the available data on each committee's deliverables and built a python tool that computes the relevant statistics and allows interactive visualization of the results.

### 4.1 Database

For each of the science committees, a simple database containing all the available data on the gender composition of its deliverables (e.g. speaker lists, promotion cases, science evaluation results, hiring shortlists) was built, spanning a time range of at least 3-5 years. The information is stored in tabular format in simple csv files, in order to be easily accessed and updated by anyone at STScI. The archival research needed to track down speaker lists for all the meetings held at STScI was particularly long and complicated, and it involved digging through old meeting posters and websites; we expect that other institutions looking to implement these guidelines might face similar hurdles. However, we find that the amount of information obtained through this process gives an overall picture that is critical to understand progress going forward, and fully justifies the effort. In order to expedite the process going forward we recommend a form per committee be created, listing all the information needed to track metrics through time. The chair of each science committee should then be in charge of filling out the form and submitting that information periodically to the appropriate office in charge of tracking the process.

Table 4 shows, as an example, the form created at STScI for the SOC of symposia and workshops. The information requested in this form expands the information currently available through data mining as the data typically available is often not enough to allow a complete and unbiased view of the process. For example, for the SOCs it is important to obtain information about the gender composition of the scientists that submitted abstracts for contributions together with the information on accepted contributions, as well.

Similarly, for committees in charge of allocating internal grants for research projects, it is important to capture information regarding the gender composition of the scientists that submitted proposals, which was not previously tracked at STScI.

**Table 4:** Example of form tailored for the SOCs. Chairs are in charge of submitting the requested information at the end of the committee's activity.

| **Committee composition:** | |
|---|---|
| Total number of participants: | |
| Number of male participants: | |
| Number of female participants: | |
| Total number of invited speakers: | |
| Number of male invited speakers: | |
| Number of female invited speakers: | |
| Total number of contributed talks (submitted, accepted): | |
| Number of contributed talks by men (submitted, accepted): | |
| Number of contributed talks by women (submitted, accepted): | |
| Total number of poster contributions (submitted, accepted): | |
| Number of poster contributions by men (submitted, accepted): | |
| Number of poster contributions by women (submitted, accepted): | |

### 4.2 Tracking Tool

The python tool loads the database, computes relevant statistics for all the science committees (e.g., fraction of men/women invited to give a talk at the symposia, fraction of men/women promoted to the next career level, etc.), and plots the results in an interactive web based format. This allows for automatic tracking once the entries in the database are updated. The WiAF is in the process of documenting the tool and is planning on sharing the software with the community in the near future.

Examples of the output of the tool are shown in Figures 1 and 2. All plots show the floor and recommended metrics (shaded areas) for the ratio of women.

Figure 1 shows the ratios of participants and people invited to give talks at the STScI symposia from 1997 to 2018. The information was not recoverable for some years. It is evident that while the ratio of female participants ranges between ~20% and ~40% (which reflects the composition through the years of the astronomy field), the ratio of female invited speakers was very biased towards low numbers in the late 90s (only 10%). In the past years, there has been a strong positive trend, bringing the ratio of female invited speakers to 30-40%.

Figure 2 shows the ratio of male (blue) and female (green) speakers invited to give a talk or assigned a contributed talk at STScI workshops between 1993 and 2017. Here as well, the information for some years is not available, nor is the information regarding invited and contributed talks. The ratio of female speakers has increased overall in the past 5 years but there are still instances in the last 5 years where the ratio of female speakers falls below the recommended floor of 30%. This demonstrates the need to establish gender guidelines for the SOC committees as well as track the outcomes on a yearly basis.

**Figure 1:** Ratio of participants (l*eft*) and people invited to give talks (*right*) at the STScI symposia divided by gender (male in blue, female in green). Not all information was available for each year. The red shaded area shows the floor (30%) while the yellow shaded area shows the recommended level for the ratio of women speakers (40%).

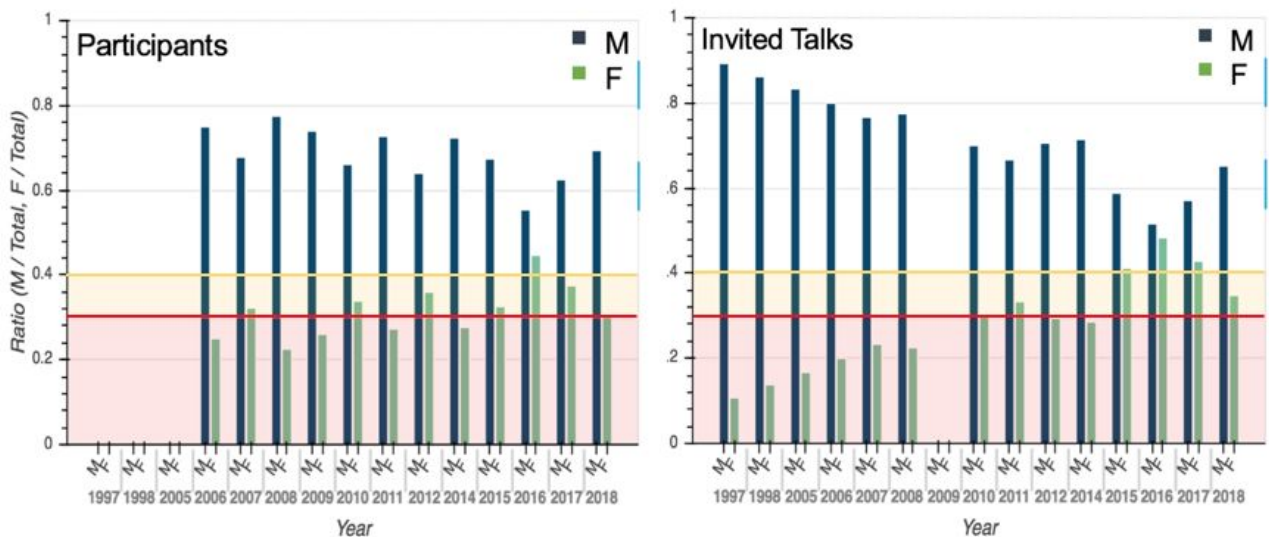

**Figure 2:** Gender composition of speakers for invited and contributed talks at STScI workshops. Not all information was available for each year. The red shaded area shows the floor (30%) while the yellow shaded area shows the recommended level for the ratio of women speakers (40%). Multiple workshops during the same year are noted with alphabet letters (e.g. 17, 17A, 17B).

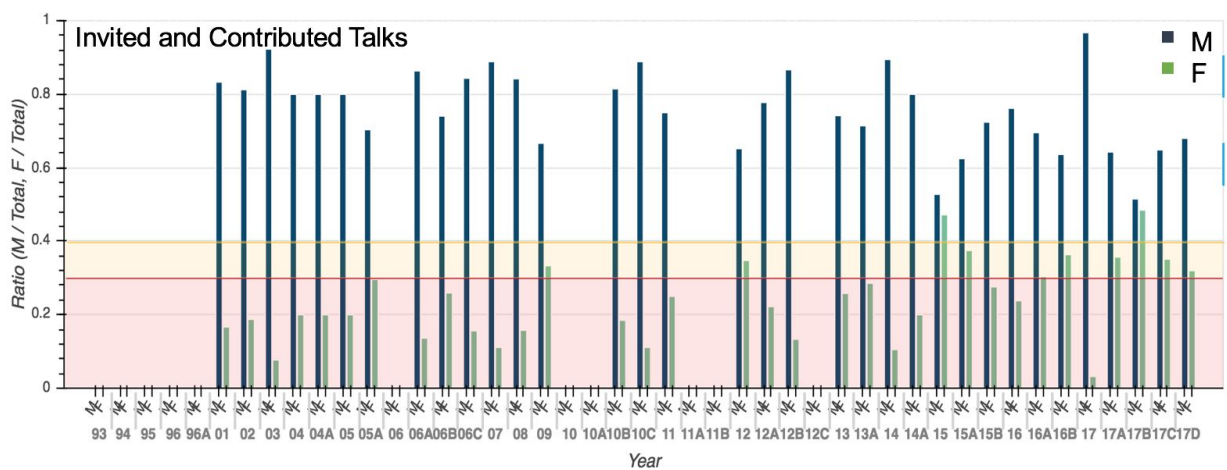

## 5. Conclusions

Similarly to other institutions, scientific committees at STScI are in charge of a variety of activities, including deliberating on the hire, renewal, and promotion of science staff members, allocating grants for science projects, evaluating the scientific productivity of the staff, and organizing a variety of science meetings. All these activities have a significant role in shaping the careers of both internal and external scientists, and need to be carried out in the most unbiased way possible. Our analysis of the baseline data for STScI shows that while there has been progress in recent years, there is still room for significant improvement in the gender composition of the scientific committees as well as in the activities they generate. While STScI has already taken several steps to increase diversity and inclusion, the guidelines and tools presented in this paper are further supporting this commitment. Currently, the guidelines focus on binary gender representation, however they can be adapted and implemented to support all minority groups.

By creating diverse committees, and making the committees aware of, and trained on implicit bias, we expect to create a diverse outcome in the activities they generate, which, in turn, will advance science further and faster. By putting in place accountability mechanisms, and by acknowledging and rewarding successful examples, we aim at establishing a virtuous cycle that will help reshaping the culture of the science staff and modify the outcomes without external enforcement.

STScI is actively working on expanding the guidelines presented in this paper to include all of STScI's committees, not just the ones related to the research staff.

A concern often raised during the discussion of the proposed guidelines is the heavier workload that women and minorities would need to take over. In this respect, an adequate recognition of this service during performance evaluation is of paramount importance in order to ensure the success of the process. STScI is also working on other possible mitigation strategies, including the creation of a list of volunteer senior women and minorities, willing to take over heavier loads in order to support more junior colleagues. It is important to note that all the WiAF members are open to the option of increasing their committee service with the final aim of increasing diversity.

**Acknowledgements**

The authors thank STScI leadership and in particular its director Kenneth Sembach, and deputy director Nancy Levenson, for fully supporting the implementation of this initiative. The authors would also like to thank Antonella Nota for fully encouraging and supporting this effort and for her valuable contributions throughout the project.